# A Novel Reconfigurable Computing Architecture for Image Signal Processing Using Circuit-Switched NoC and Synchronous Dataflow Model

Feitian Li, Fei Qiao, Qi Wei, Huazhong Yang[1]
[1]National Laboratory for Information Science and Technology,
Institute of Circuits and Systems,
Dept. of Electronic Engineering, Tsinghua University, Beijing 100084, P.R. China
E-mail: lft10@mails.tsinghua.edu.cn, {qiaofei, weiqi, yanghz}@tsinghua.edu.cn

**Abstract**

In this paper, a novel reconfigurable architecture is proposed for multifunctional image signal processing systems. A circuit-switched NoC is used to provide interconnection because the non-TMD links ensure fixed throughput, which is a desirable behavior for computational intensive image processing algorithms compared with packet-switched NoC. Image processing algorithms are modeled as synchronous dataflow graphs which provide a unified model for general computing procedure. An image processing system is considered as several temporally mutually exclusive algorithms. Thus, their dataflow graph representations could be considered as a group and a merging algorithm could be applied to generate a union graph while eliminating spatial redundancy for area consumption optimization. After the union graph have been mapped and routed on the NoC, the reconfigurable system could be configured to any of its target image processing algorithms by properly setting the NoC topology. Experiments show the demo reconfigurable system with two image processing applications cost 26.4% less hardware resource, compared with the non-reconfigurable implementations.

## 1. Introduction

The rapid growing of contemporary image processing algorithms has been presenting essential challenges on modern integrated circuit technology as well as designing methodologies. In this paper an efficient reconfigurable architecture is discussed as a possible solution for large scale multifunctional image processing computing paradigm which guarantees high performance and encourages model reuse which reduces area consumption as a result.

NoC facilitates the implementation of reconfigurable computing systems by proving a time-variant, scalable interconnection paradigm. A circuit-switched NoC's link is occupied exclusively by a connection between two routers, which differs from a packet-switched NoC which routed each of its packet according to forward tables in routers and address information in packet headers. The main reason why circuit-switched NoCs are favorable for image signal processing reconfigurable systems is that throughput on each link is fixed as the bandwidth, which corresponds to the image processing algorithm's stable computation load and high performance requirement. Additionally, since no extra routing information is needed, a circuit-switched NoC could reach maximum data transmitting efficiency which helps reduce power consumption.

The computing procedure of algorithms could be described as a dataflow graph with its edges representing the data paths and nodes the operators [1][2]. In the dataflow graph model, computation is seen as data tokens passing between operators (nodes) directed by the directed edges. As discussed in future chapters, an algorithm could be implemented on a circuit-switched NoC with each of the nodes in its dataflow graph representation assigned to a router. Upon proper configuration of the NoC topology, edges in the graph are formed by NoC connections, and the algorithm is activated.

This paper is organized as followed: the general architecture of our system is briefly addressed, key concepts is examined comprehensively, detailed design flow as well as a simple experiment is presented, and a conclusion is made finally.

## 2. General Architecture

The adaption of NoC and dataflow graph suggests a straight-forward overall architecture of reconfigurable system. A circuit-switched NoC provides connections which serve as the directed edges in a dataflow graph, and each input or output port of a node in a dataflow graph is assigned to a router in the NoC.

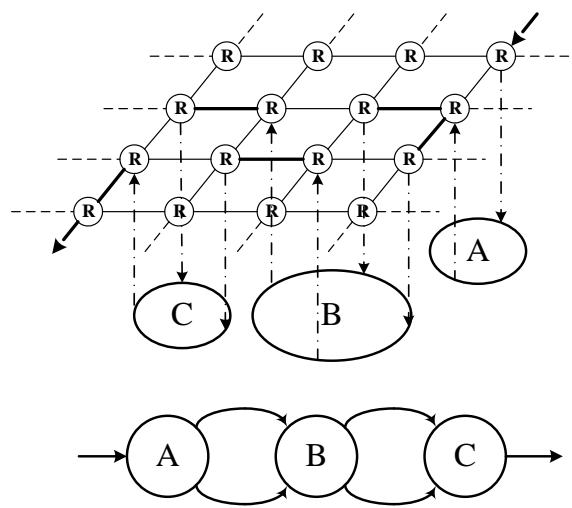

**Figure 1:** An illustrative application on our system architecture (NoC topology is 2D mesh) and its dataflow representation (circles labeled "R" stand for routers; bold connections between routers are switched on; A, B and C are nodes)

A simple router comprises the NoC in this paper which could be modeled as a crossbar with a configuration model is used. The configuration model can perform a reconfiguration to the crossbar to turn it into any possible status upon an external configuration signal [3]. The connection to the router's four neighbors only allows data to go through it in one direction at a time spot and the port connected to the local port could only be configured as input or output at a time.

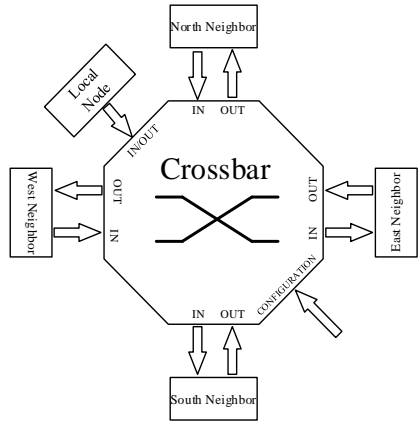

**Figure 2:** Router schematic

A computing system might apply different algorithms to its inputs at different time. Thus, a system is modeled as a group of temporally exclusive algorithms which could be modeled as a set of directed graph of different topologies. Switching between algorithms suggests an alternation of dataflow graph. Thus, to provide proper configuration information of the NoC, a merging algorithm is needed to combine individual graphs to a union graph which includes all nodes and edges. After mapped and routed to a circuit-switched NoC, each of these algorithms could be selected and activated by reconfiguration of the NoC.

## 3. Circuit-Switched NoC and Synchronous Dataflow Graph Revisited

### 3.1. Circuit-Switched Network-on-Chip

A circuit-switched NoC could be modeled as an graph $N = (R, L)$, with $R$ representing the set of local ports of routers and $C$ the set of links. Notice that $C$ contains the links both inside the crossbars of routers and between routers.

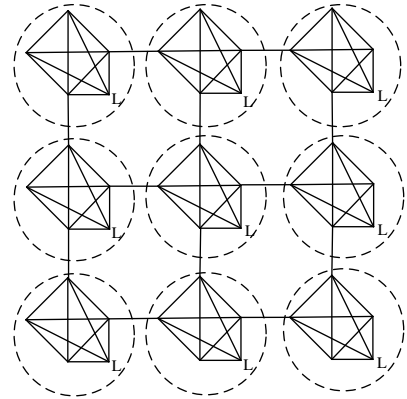

**Figure 4:** The topology of a 3X3 2D mesh NoC (each router is enclosed by a circle of dotted line, vertexes labeled "L" represents local ports), each router could be seen as a 5X5 crossbar.

To a circuit-switched NoC, the main challenge is the map and route problem [4][5] :

Given a graph $G = (V, E)$ in which an edge is donated as $e = (v_d; v_{l1}, v_{l2}, \ldots, v_{ln}; w)$ with a single driver $v_d$, possible multiple loads $v_{l1} \sim v_{ln}$ and an index $w$, determine the map function

$$\sigma_M : V \to R$$

and the route function

$$\sigma_R : E \to 2^L$$

such that:
1. $\sigma_M$ is injective.
2. $\sigma_R(e)$ is a connected subnet and includes $\sigma_M(v_d)$ and every $\sigma_M(v_{li})$ as its vertexes.
3. $\sigma_R(e_1) \cap \sigma_R(e_2) = \emptyset$ if $e_1$ and $e_2$ are indexed by same $w$.

The circuit switched property of the NoC is implied by the third constraint that two connections presenting on the NoC at the same time should not share any links. A special case of this problem with a unified $w$ have been considered extensively in traditional VLSI physical design methodologies [6][7].

### 3.2. Synchronous Dataflow Graph

Dataflow graphs have been used extensively in the conventions of digital signal processing [1]. They are frequently used to model data processing in computing systems.

In dataflow graphs, a computation is recognized as data tokens passing through a set of operators in a pre-defined order. The nodes in the graphs represents processing operators, and the edges data paths. As a node might have more than one input or output ports, the term "vertex" is used to donate these ports. In this paper, a dataflow graph could be described as $G = (V, E)$ or $G = (N, E)$.

There are certain basic constraints to behavior of nodes:
- Casualty. A node might take arbitrary input tokens, but its outputs must be determined by its initial internal states (if any) and all its inputs till now.

- Locality. A node must apply its own computational laws with inputs and internal states as its only reference. To paraphrase, information is not allowed to be passed among nodes.
- Fixed input and output. A node must consume a fixed amount of inputs and must generate a fixed amount of outputs if it wants to.

A node might take a set of tokens from its input ports simultaneously, and generate the corresponding output after some time delay. Such behavior is called "firing" of a node [1]. A node is allowed to fire for infinite times. In this work, a most natural way of firing is considered: a node fires upon an external signal informing that the input is ready, and it must fire correctly each time the signal is set. Behavior like this is referred to as "self-timed firing" [1][2].

Another important consideration on constraints of nodes is timing. To avoid unpredictable behavior, all nodes must fire synchronously. That is, the sample time spot of all nodes must be exactly the same [2]. However, the definition of synchronous dataflow graph is still inconvenient, and the last constraint mentioned above shall be replaced by a stronger one:

- Guaranteed outputs. A node must provide a fixed amount of output tokens for every set of inputs after some time delay.

The constraint seems serious, however there is no essential difference compared to the general ones: a node could define a null token indicating that no information is contained is this output token, and a node could resynchronize all its input streams with no index information overhead.

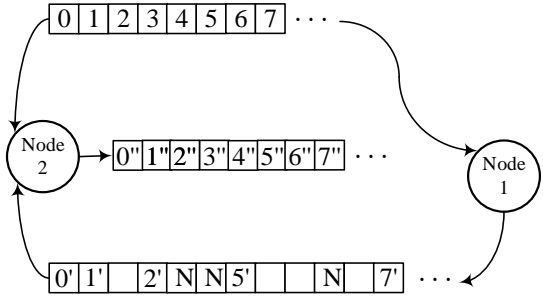

**Figure 3:** Streams with null token (tokens with mark "N" in this figure) and delay (tokens with no mark) could be resynchronized.

Under all these constraints, a node could fire infinitely if there is an endless stream of input is supplied, generating an endless stream of output stream. Notice that each token in input and output stream is matched, so a node is safe to assume that it reads the same set of input tokens if the start of input streams is correctly synchronized and no extra index information is needed.

In the context of NoC as link between nodes, the issue of time delay must be dealt carefully. Since a dataflow graph representing an abstract computational process should not make any assumption about the NoC that provides connectivity (edges) of the graph, the delay caused by router must be considered unknown and may be present on any of the links.

Directed acyclic graphs are immune to unknown delays on its edges, which means streams on each edge could be generated correctly and nodes are guaranteed to function correctly if it resynchronizes all its input streams.

However, this does not means that algorithms represented by cyclic graphs could not be realized. Cyclic graphs could be acyclic under proper division. Additionally, notice that the nodes attached to NoC routers are not necessarily corresponding to those in dataflow graphs. In fact, each NoC node could be an arbitrary division of the original directed acyclic dataflow graph with some arcs connected internally (i.e., not by NoC), which in effect provides area optimization space.

## 4. Proposed Design Flow and Experiments

The general architecture of the proposed system decouples the description or potential coding work of applications from the mapping and routing task of the NoC. Therefore, design flow could be broken down into several steps and only one central problem need to be considered at each step.

The outline of a possible design flow is as followed:
1. Description of individual algorithms.
2. Optimization of individual algorithms.
3. Merging operation of individual application graphs. An algorithm is needed to summarize the interconnection information scattered in individual graphs to generate a union graph which contains complete link information to guide the reconfiguration of NoC. Recognition and elimination of spatial redundancy is also done in this stage.
4. Mapping and routing. Mapping and routing is a procedure to determine for each input and output port of every node in the union graph which router should be assigned, and which routers should provide intermediate hops for NoC links.

Each step will be explained in details.

### 4.1. Algorithm Description and Optimization

In this stage, a set of desirable multimedia (or any application satisfying the profile we give in introduction) applications is chosen. The desired function is that each of these applications (algorithm) could be activated and switched to another at any given time.

Each application is described by a directed acyclic dataflow graph (most probably generated automatically from compilers or synthesizers, and cyclic semantics must be packed to form an acyclic graph after division). Some local optimizations could be applied to eliminate redundancies introduced by the possible use of high-level descriptive language. This problem is known as basic block optimization [8] in the conventions of compiler design.

### 4.2. Global Redundancy Detection

Let $G_1, G_2, \ldots, G_N$ be the individual graphs generated by the previous stage. Now interconnection information distributed in each graph must be assembled into a union graph $G = (N, E)$ to provide complete map and route requirements. Global redundancies elimination is also done in this stage. There might be several kinds of redundancies, in our work we focus on area redundancy, or the optimization of area consumption. To address the problem formally,

Given $G_1, G_2, \ldots, G_N$, try to minimize the metric
$$AREA = \sum_N AREA(n) + A_{\text{routing}}$$

Since the area consumption associated with routing could not be known at this stage (and it should not be coupled with any specific NoC structure or map and route algorithm), we omit it and expand the summation:

$$\sum_N AREA(n) = \sum_N (AREA_{\text{int}}(n) + AREA_{\text{router}})$$

where $AREA_{\text{int}}(n)$ is the intrinsic area consumption of the node cost by its signal processing procedure and $AREA_{\text{router}}$ is the area consumption of a router.

A conclusion could be drawn that area optimization is roughly equivalent to the minimization of node count in the union graph. Hence, a straightforward heuristic algorithm could be expressed as followed:

1. Start. Label the nodes in $G_1, G_2, \ldots, G_N$. Each node shall have the form $P_m^n$ with $P$ indicating the node's type (e.g., adder, multiplier, FFT), $n$ the graph where it occurs, and $m$ the current count of its appearance in a graph (e.g., the third appearance of an adder in the second application graph should be represented as $\text{ADDER}_3^2$).
2. The nodes of the union graph is determined. For every label type $P$, it shall prepare $\max_{\text{in all } P_m^n} m$ copies. A function
$$\sigma_N: \bigcup_{i=1}^N N(G_i) \to N(G)$$
could be determined, specifying every application graph's node's corresponding node in $G$.
3. Determine the edges. If there is a connection from $v_1$ to $v_2$ in $G_i$, then use a color $i$ to draw a edge connecting $\sigma_V(v_1)$ to $\sigma_V(v_2)$ in $G$. Another function
$$\sigma_E: \bigcup_{i=1}^N E(G_i) \to E(G)$$
could also be determined, indicating every application graph's edge's corresponding edge in $G$.
4. Set $q = 0$.
5. Packing. Choose randomly a color combination in $G$. If at least one edge in this color combination has unmarked node(s), mark the unmarked node(s) with number $q$, observe if one of their other edges is also in this color combination. If so, mark these edge's other nodes with number $q$ if these nodes do not have a mark. Keep the color combination unchanged, repeat doing observation and marking until no edges could be found. Increase $q$ by 1.
6. Repeat step 5 until all nodes are marked.
7. Division. Make a division so that all nodes in $G$ with the same mark is packed together. Quit.

The following figures give an example of how this algorithm actually works.

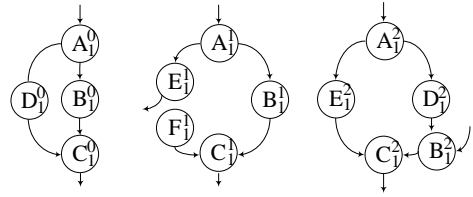

**Figure 5a:** Marking of application graphs. (step 1)

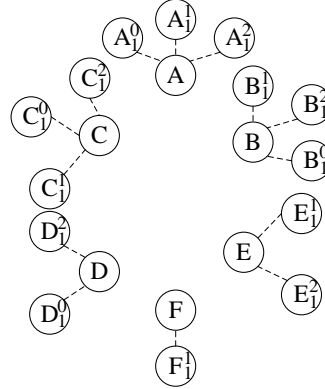

**Figure 5b:** Nodes of union graph and their relation ($\sigma_V$) to the nodes in figure 5. (step 2)

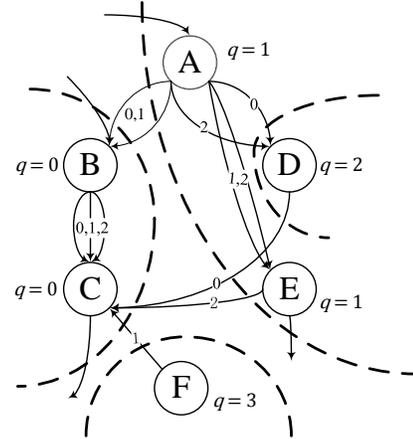

**Figure 5c:** The numbers on the edges corresponds to the colors in the algorithm description. Nodes enclosed in bold dotted lines are packed. (step 3~6)

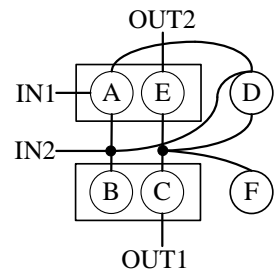

**Figure 5d:** Final result after division (step 7).

### 4.3. Map and Route

The main problem of map and route has been addressed in the previous chapter. Notice that the edge set $E$ in $G$ (probably generated by the algorithm in the previous section) corresponds to $L$ in the map and route problem. For applications with a comprehensive power model, power consumption might be estimated after the map and route algorithm gives the exact routing of the system. Map and route problems have been studied for FPGAs [6] in which a similar problem is addressed, a set of algorithms have been proposed and studied, many of them are already in industrial fields.

### 4.4. Operation of System

Proper operation of the system is quite straightforward. If an algorithm represented by $G_i = (V_i, E_i)$ is to be activated, switch on the links on $\sigma_R(\sigma_E(E_i))$, and switch off the other links.

To give the system a friendly behavior, additional stream start and end behavior might be defined for each node. Configuration information of NoC is determined after the proposed design flow is completed and could be given by components outside the system.

### 4.5. Experiment

An illustrative experiment has been done to prove the feasibility of the proposed architecture. Consider the algorithms for a processor working in an unpredictable illuminant condition. In circumstances of strong illuminant (day), color calibration [11] is required to balance the energy distribution between RGB channels to magnify subtle details of the frame which would otherwise be omitted by the edge detection algorithm's threshold; whereas in a dark surrounding (night), histogram equalization [12] is crucial to enhance the gradient of the frame color field. In both condition, an anti-noise Gaussian filter [12] is used to smooth the frame before other procedure, and a canny [10][12] operator extracts the edges from the frame in the end. Hence, the system could be represented by the following schematic,

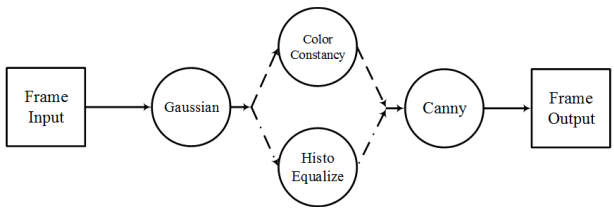

**Figure 6:** Experiment system schematic.

After map and route, a 2X5 mesh NoC is used to provide interconnection.

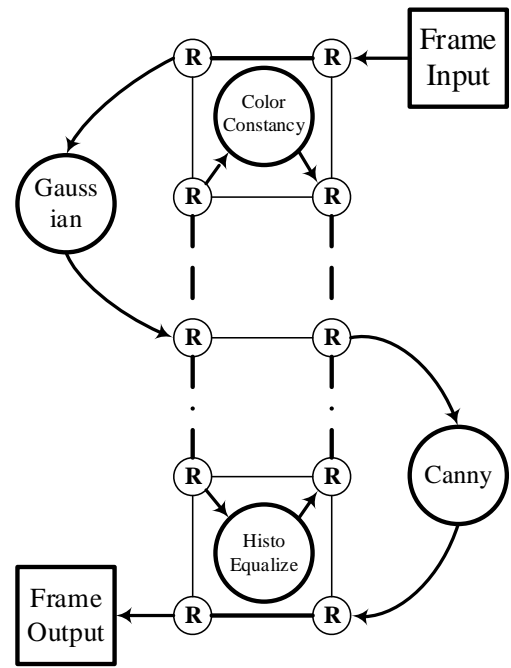

**Figure 7:** NoC schematic, router links drawn in bold is switched on conditionally to form a time-variant interconnection (single channel).

This experimental system is implemented on a XC6VLX365T Virtex 6 Xilinx FPGA. Model's resource utilization is shown in the following table:

**Table 1:** Resource utilization of models

| Module | Gaussian Filter (Single Channel) | Color Constancy (3 Channel) | Histogram Equalization (Single Channel) | Canny Operator (Single Channel) |
|---|---|---|---|---|
| Slices Occupied | 1058 | 460 | 5503 | 3843 |

Comparison of the resource utilization and performance of the entire preprocessor system is summarized in the following table.

**Table 2:** Preprocessor resource consumption and performance comparison

| Preprocessor System (3 Channel) | Day | Night | Reconfigurable (with NoC) |
|---|---|---|---|
| Slices Occupied | 14327 | 27872 | 31042 |
| Maximum Frequency (MHz) | 37.071 | 65.504 | 37.071 |

The reconfigurable system has a resource cost (14327+27872-31046) / (14327+27872) = 26.4% lower than two individual functionally equivalent preprocessors. The timing performance is not affected by the reconfigurable architecture since the structure of circuit-switched NoC routers is simple compared to NoC nodes and would not be the bottleneck of maximum clock frequency.

### 5. Conclusion

In this paper, a novel architecture combining circuit-switched NoC and dataflow graph is discussed. The main consideration of choosing this combination rather than the paradigm of some other works ([13][14]), in which multicores are interconnected by packet-switched NoCs, is that performance should be the most significant factor for image processing algorithms, so that operators must be implemented by hardware approaches and the overhead of router hops must be trivial. Getting down to the proposed architecture, the synchronous dataflow is used to provide a simple and unified model for general computing procedures which facilitates the design flow and potential optimization methods, and nodes in dataflow graphs could be implemented directly as nodes in the NoC. The simplicity of circuit-switched NoC router ensures that it would not be a bottleneck of overall performance, which makes the implementation of operators the decisive factor of performance.

The merging algorithms in chapter 4 exploits the idea of area cost optimization through recognizing of same operator among different graphs and packing of subnets. However, a more practical problem is to define the proper granularity of nodes in a dataflow graph. Empirical approaches might be useful with a knowledge of conventional image processing algorithms, but a more reasonable solution is to break down each algorithm into the most basic semantics so that the merging algorithm might detect redundancies to the most extent. Additionally, large scale systems might contain complex algorithms. A plausible solution to build such system is to use languages similar to contemporary computer programming languages. Modern compilers adopted the idea of dataflow graph analysis to fragments of program as one of its optimization tools and intermediate representation of semantic [8][9]. It is reasonable to expect that compilers could take a step further and perform proper transformations to accommodate NoC systems. Both the requirement to detect redundancies and the need to build complex systems suggest an architecture-specific compiler could be used as a front end to the proposed design flow, which could be of interest for our future works.

## 9. References

[1] Lee, Edward A., and David G. Messerschmitt, "Synchronous data flow", Proceedings of the IEEE 75.9 (1987): 1235-1245.

[2] Ghamarian, Amir Hossein, et al, "Throughput analysis of synchronous data flow graphs", Application of Concurrency to System Design, 2006. ACSD 2006. Sixth International Conference on. IEEE, 2006.

[3] Wolkotte, Pascal T., et al, "An energy-efficient reconfigurable circuit-switched network-on-chip", Parallel and Distributed Processing Symposium, 2005. Proceedings. 19th IEEE International. IEEE, 2005.

[4] Hansson, Andreas, Kees Goossens, and Andrei Rădulescu, "A unified approach to constrained mapping and routing on network-on-chip architectures", Proceedings of the 3rd IEEE/ACM/IFIP international conference on Hardware/software codesign and system synthesis. ACM, 2005.

[5] Li, Ming, Qing-An Zeng, and Wen-Ben Jone, "DyXY: a proximity congestion-aware deadlock-free dynamic routing method for network on chip", Proceedings of the 43rd annual Design Automation Conference. ACM, 2006.

[6] Sherwani N A, Algorithms for VLSI physical design automation, Kluwer Academic Publishers, 1995.

[7] Sait S M, Youssef H, VLSI physical design automation: theory and practice, World Scientific, 1999.

[8] Aho A V, Compilers: principles, techniques, & tools, Pearson Education, India, 2007.

[9] Fischer C N, Cytron R K, LeBlanc R J, Crafting a compiler, Addison-Wesley Publishing Company, 2009.

[10] Canny, John, "A computational approach to edge detection", Pattern Analysis and Machine Intelligence, IEEE Transactions on 6 (1986): 679-698.

[11] Ebner M, Color constancy, Wiley. com, 2007.

[12] Jain, Ramesh, Rangachar Kasturi, and Brian G. Schunck, Machine vision. Vol. 5. New York: McGraw-Hill, 1995.

[13] Avakian, Annie, et al, "A reconfigurable architecture for multicore systems", Parallel & Distributed Processing, Workshops and Phd Forum (IPDPSW), 2010 IEEE International Symposium on. IEEE, 2010.

[14] Fan, Hongbing, Yue-Ang Chen, and Yu-Liang Wu, "R-NoC: an efficient packet-switched reconfigurable networks-on-chip", Reconfigurable Computing: Architectures, Tools and Applications. Springer Berlin Heidelberg, 2012. 365-371.